\documentclass[sigconf,screen,authorversion]{acmart}

\usepackage[utf8]{inputenc}    
\usepackage{booktabs,multirow} 

\newcommand{\shiftup}{\vspace{-0.5em}}

\setcopyright{acmlicensed}
\copyrightyear{2018}
\acmYear{2018} 
\acmPrice{15.00}
\acmDOI{10.1145/3197091.3197123}
\acmISBN{978-1-4503-5707-4/18/07}
\acmConference[ITiCSE'18]{23rd Annual ACM Conference on Innovation and Technology in Computer Science Education}{July 2--4, 2018}{Larnaca, Cyprus}

\begin{document}

\title{Enhancing Cybersecurity Skills by Creating Serious Games}

\author{Valdemar Švábenský}
\orcid{0000-0001-8546-280X}
\affiliation{
  \institution{Masaryk University}
  \department{Faculty of Informatics}
}
\affiliation{
  \department{Institute of Computer Science}
  \streetaddress{Botanická 68a}
  \city{Brno} 
  \country{Czech Republic}
  \postcode{60200}
}
\email{svabensky@ics.muni.cz}

\author{Jan Vykopal}
\orcid{0000-0002-3425-0951}
\affiliation{
  \institution{Masaryk University}
  \department{Institute of Computer Science}
  \streetaddress{Botanická 68a}
  \city{Brno} 
  \country{Czech Republic}
  \postcode{60200}
}
\email{vykopal@ics.muni.cz}

\author{Milan Cermak}
\orcid{0000-0002-0212-6593}
\affiliation{
  \institution{Masaryk University}
  \department{Faculty of Informatics}
}
\affiliation{
  \department{Institute of Computer Science}
  \streetaddress{Botanická 68a}
  \city{Brno} 
  \country{Czech Republic}
  \postcode{60200}
}
\email{cermak@ics.muni.cz}

\author{Martin Laštovička}
\orcid{0000-0002-6604-6947}
\affiliation{
  \institution{Masaryk University}
  \department{Faculty of Informatics}
  \department{Institute of Computer Science}
  \streetaddress{Botanická 68a}
  \city{Brno} 
  \country{Czech Republic}
  \postcode{60200}
}
\email{lastovicka@ics.muni.cz}

\begin{abstract}
Adversary thinking is an essential skill for cybersecurity experts, enabling them to understand cyber attacks and set up effective defenses. While this skill is commonly exercised by Capture the Flag games and hands-on activities, we complement these approaches with a key innovation: undergraduate students learn methods of network attack and defense by creating educational games in a cyber range. In this paper, we present the design of two courses, instruction and assessment techniques, as well as our observations over the last three semesters. The students report they had a unique opportunity to deeply understand the topic and practice their soft skills, as they presented their results at a faculty open day event. Their peers, who played the created games, rated the quality and educational value of the games overwhelmingly positively. Moreover, the open day raised awareness about cybersecurity and research and development in this field at our faculty. We believe that sharing our teaching experience will be valuable for instructors planning to introduce active learning of cybersecurity and adversary thinking.
\end{abstract}



\maketitle

\section{Introduction}

As the importance of securing computer systems grows, so is the cybersecurity workforce shortage. It is expected that by 2022, 1.8~million jobs that require cybersecurity expertise will be unfilled~\cite{jtf-csec}, stressing the importance of educating security professionals. An efficient security education balances theoretical knowledge and concepts, traditionally taught at universities, with practical applications in real-world settings. 

To complement traditional courses on computer networks and security at our faculty, we developed two practical, project-oriented courses on cyber attacks and defense. We focus especially on adversary thinking, a crucial skill for cybersecurity experts who must be able to think like an attacker in order to set up effective countermeasures. While this skill can be exercised in Capture the Flag games, challenges, and competitions, our courses introduce an innovative approach. The learners are guided to create a serious security game deployed at the KYPO cyber range~\cite{kypo}, which allows emulating real threats and attacks in a controlled environment.

With respect to the thorough literature survey below, we claim that our courses are unique in combining a serious game project with hands-on cybersecurity. In this paper, which can be of particular interest to fellow security instructors, we share a detailed description of the design, content, and assessment methods of the courses. Next, we present and analyze student surveys. Finally, we share the lessons learned in three semesters of teaching and continuously innovating the courses. We highlight that the students' projects transfer to practice. All learners present their results to other students of the faculty at an open day event, and the best games are even used for further training by our security team.

\section{Related work} \label{sec:related}

To map the current landscape of similar courses, we searched for existing university courses and literature as well as curricular guidelines for cybersecurity education. We examined course catalogs of the 10 currently top-ranked computer science (CS) universities based on research
and teaching
(as listed in QS and THE World University Rankings, respectively). In the course catalog of each university, we searched for topics such as (cyber)security, networks, or cryptography. The search was restricted to the CS and engineering departments. We discovered relevant courses, for example, MIT's Network and Computer Security\footnote{\url{https://ocw.mit.edu/courses/electrical-engineering-and-computer-science/6-857-network-and-computer-security-spring-2014/index.htm}}, dealing with cryptography, secure programming, or internet security. Most often, the courses were taught using a combination of lectures, (group) homework assignments, (group) projects, student presentations, and a final test. However, we did not find any course dealing in-depth with penetration testing, learning by teaching, and creating serious games in a cyber range as our courses do.

We also examined related papers from the past 5 years on major CS education conferences such as ACM SIGCSE, ITiCSE, ICER, and USENIX ASE (formerly 3GSE). The publications describe learning cybersecurity skills in a practical and engaging way by using video games~\cite{Thompson}, Capture the Flag events~\cite{Taylor}, and even card games~\cite{Denning}. Gamification of cybersecurity courses by adding a background story~\cite{chothia}, competitive game elements~\cite{Dabrowski}, or experience points~\cite{Schreuders} is popular, resulting in increased student motivation, interest in security topics, and performance. While hands-on cybersecurity labs described in literature~\cite{OLeary,Timchenko} cover similar content as our courses, they do not include a serious game project as a teaching method. On the other hand, existing game development courses~\cite{Yun,krusche} focus mostly on programming or game design as such but not on cybersecurity.

\section{Description of the courses} \label{sec:courses}

This section describes the current model of our two courses, the Introductory and the Follow-up. We established the model after three semesters of experience. The courses are offered for CS university students and include 12 weeks of 2-hour sessions plus homework assignments. Table~\ref{table:outcomes} lists the learning outcomes of both courses\footnote{Selected learning outcomes correspond to the following NIST NICE~\cite{nist-nice} knowledge, skills, and abilities descriptions: K0003, K0005, K0013, K0070, K0106, K0177.}. Table \ref{table:course-structure} provides the schedule and the structure of the courses.

The courses follow recent standards and comprehensive curricular guidelines for cybersecurity education, namely these NSA/DHS CAE Knowledge Units~\cite{nsa-ku} and corresponding NIST NICE competencies~\cite{nist-nice}: cyber defense, cyber threats, networking concepts, network defense, and penetration testing. Moreover, the courses strongly correspond with NCC Information Security Curricula~\cite{ncc} courses Network Security I and Ethical Hacking. Finally, they cover selected topics of the following knowledge units in the Joint Task Force Cybersecurity Curricula~\cite{jtf-csec}: Data integrity and authentication, Network Defense, and System Control, along with adding extra features described in this section.

Apart from connecting to the cybersecurity education guidelines, the courses also employ methods grounded in pedagogical theory and teaching practice~\cite{petty}. The sessions include a combination of lectures, supervised student practice, and group work. Using the terminology of an extensive survey of active learning methods in computer science~\cite{activelearning}, we focus on \textit{integrating labs with lectures}, \textit{cooperative learning}, and \textit{project-based learning}.

\begin{table*}
\shiftup
\renewcommand\arraystretch{1.2}
\begin{center}
\caption{The learning outcomes of both the Introductory and the Follow-up course\shiftup}
\begin{tabular}{|l|p{10cm}|p{5.1cm}|}
\hline
& \textbf{Cyber Attack Simulation (Introductory)} & \textbf{Cyber Defense Tutorial (Follow-up)} \\ \hline
\textbf{Knowledge} & Describe the stages of a cyber attack & \\ \cline{2-2}
& Understand system and application security threats and vulnerabilities (e.\,g., authentication attacks, DoS attacks, MitM attack, OWASP Top 10 vulnerabilities) & \\ \cline{2-2}
& Name cyber defense and vulnerability assessment tools and their capabilities & \\ \cline{2-2}
& Explain laws, regulations, policies, and ethics related to security and privacy & \\ \hline
\textbf{Skills} & Perform penetration testing focused on a particular threat or vulnerability & Secure a particular network service or application (e.\,g., Apache or Wordpress) \\ \cline{2-3}
& Use a cyber range both as a learner and as a designer of games running in it & Perform penetration testing of the service or application \\ \hline
\textbf{Experience} & Give a presentation explaining the vulnerability selected for the game & \\ \cline{2-3}
& \multicolumn{2}{|p{15.1cm}|}{Practical work in small teams (Introductory) or individual (Follow-up) including setting up and maintaining systems, assessing their vulnerabilities, and developing a new serious game or a gamified training tutorial}  \\ \cline{2-3}
& \multicolumn{2}{|p{15.1cm}|}{Give two presentations of the final project (Test run and Open day) and instruct learners who use it} \\ \hline
\end{tabular}
\label{table:outcomes}
\end{center}
\end{table*}

\begin{table*}
\shiftup
\renewcommand\arraystretch{1.2}
\begin{center}
\caption{The schedule and the structure of both the Introductory and the Follow-up course\shiftup}
\begin{tabular}{|l|p{1.3cm}|l|l|l|l|l|l|l|l|l|l|l|l|}
\hline
\textbf{Week}     & 1 & 2\phantom{...} & 3\phantom{...} & 4 & 5 & 6 & 7 & 8 & 9 & 10 & 11 & 12 & 18 \\ \hline
\textbf{Introductory}    & Exemplary game & \multicolumn{5}{|p{3.4cm}|}{Network security basics, hands-on labs, homework} & \multicolumn{1}{|p{2.8cm}|}{Game design tutorial, topic choice} & \multicolumn{3}{|p{1.8cm}|}{Presentations, consultations} & \multirow{4}{*}{\shortstack[l]{Test\\run}} & \multirow{4}{*}{\shortstack[l]{Open\\day}} & \multirow{4}{*}{\shortstack[l]{Final\\result}} \\ \cline{1-11} 
\textbf{Follow-up} & Topic choice & \multicolumn{3}{|p{1.8cm}|}{Concept consultations} & \multicolumn{1}{|p{1.8cm}|}{Concept finalization} & \multicolumn{4}{|l|}{Technical consultations} & Presentation & & & \\ \hline
\end{tabular}
\label{table:course-structure}
\end{center}
\shiftup
\end{table*}

\subsection{Cyber Attack Simulation (Introductory)}
The first course focuses on the basics of offensive cybersecurity. It provides both theoretical and practical experience to students, who elaborate a game project on penetration testing in teams of two people (three if the total number of students is odd). The course is intended for at most 18 undergraduates (sophomores and juniors) who passed prerequisite courses on privacy and computer networks and systems, can read technical papers, and write in English.

In the first session, students play an exemplary security game to see what we expect from their project. During the first, theory part of the course, lectures and exercises introduce students to security topics. The acquired knowledge and skills are used in the second, project part of the course during the creation of Capture the Flag games, whose form is described in~\cite{SIGCSE-paper}. The course ends with the Open Day event, where the students publicly present their games.
 
The theory part of the course explains cybersecurity topics based on lecturers’ practical experiences from the day-to-day operation of a university computer security incident response team (CSIRT). Students get familiar with common attacks (such as scanning, exploits, and threats specified by the OWASP Top 10 project), basics of forensic analysis, and attack detection and mitigation. During the sessions, students perform hands-on exercises to immediately verify their theoretical knowledge in practice. Each student has own learning sandbox in the KYPO cyber range containing attacking and vulnerable machines. The sessions are followed by weekly homework assignments, which involve using serious games created in the previous course runs, tasks in the learning sandbox, or freely available online resources\footnote{Such as \url{https://www.hackthissite.org} or \url{https://hack.me/}.}.

During the project part of the course, student teams create their own game. The instructors form the teams based on students' performance in the exemplary game and a subsequent survey. The aim is to pair less skilled or less experienced individuals with advanced learners to simulate real workplace. Afterward, students are introduced to the basics of creating serious games. Then, they select a game topic, such as exploiting a vulnerability to gain access to a system, from a list of predefined topics. Each team proposes a story of the game, designs its levels, and prepares virtual machines in the cyber range on which the game will be played. During the sessions, each team presents the proposal to other students and lecturers and receives feedback to improve the proposal. The results are tested by members of the CSIRT, allowing the teams to find out how experts react and how the game can be adjusted. The entire course is concluded by the Open Day event where the visitors (usually students from the whole faculty) play the created games under the guidance of the student authors. Based on the feedback received at the event, the students finish the games including supplementary materials and submit them to instructors for the final evaluation.

Although the course is taught in the Czech language, all materials used by instructors or prepared by students (for both the presentations and projects) are in English. However, the instruction itself and communication within students' teams is conducted in Czech. This is motivated by the fact that English is considered a common language in the cybersecurity field, but also by the concern that communication only in English might create a barrier for those less fluent in the language.

\subsection{Cyber Defense Tutorial (Follow-up)} \label{subsec:followup}
The second course is offered for at most 6 students who passed the Introductory course. This number of students enables the instructor to advise student projects thoroughly.

The students learn how to secure a particular network service or application by designing a gamified tutorial on that topic. The tutorial consists of step-by-step instructions that enable the learner to secure the service or application running on a host in the cyber range. This part may be followed by automated attacks against the service or application, immediately enabling learners to test whether their countermeasures were set properly. Students are allowed to enroll in the course repeatedly in multiple semesters if they choose different project topic each time.

In contrast to the Introductory course, students work on their projects individually and from the beginning, without the theoretical introduction by instructors. The whole course is driven by students rather than teachers. Since the students completed the Introductory course, they already have experience in creating a game in the cyber range. This allows them to focus on the project topic chosen in the first week. In the following weeks, they elaborate the game outline on their own at home and consult ideas and issues they encounter with the instructor at course sessions. Besides the consultations, the sessions contain brief lectures on game tutorial creation essentials, including automation of installation and host configuration in the cyber range or orchestration of attacks execution. The students also advise the learners from the Introductory course, since we believe that advice from peers has a larger impact on the students compared to the instructors. Similarly as in the first course, the students present and discuss their projects to get feedback not only from the instructor but also from their classmates. The test run and the Open Day is shared with the Introductory course, followed by a final submission 6 weeks afterward.

\shiftup
\section{Project assessment methods} \label{sec:assessment}

\subsection{Formative Assessment}
While working on their game, students receive formative feedback in three settings: presentations of project milestones to the class, consultation sessions with the course tutors, and a test run of the game with security experts. All three occasions are detailed below.

\subsubsection{In-class presentations}
Both courses require the students to present their progress in brief talks. Students are given the structure of the presentation with the aim to help them focus on the content (e.\,g., explaining details of the vulnerability). The time limit of 5 to 10 minutes should force them to prioritize the key messages. The talks provide opportunities to receive feedback, both on the content and the presentation delivery, not only from instructors but also from peers. This is beneficial for the presenting team and even for other classmates observing the comments.

\subsubsection{In-class consultations}
In-person or e-mail consultations of students' projects is a prevalent method of the formative assessment, especially in the Follow-up course. Although students are instructed to find answers to their questions within the class or from open sources first, they are encouraged to ask the instructors for help with the cyber range. Learners often struggle with the configuration of virtual hosts, the specifics of a cyber range and the underlying virtualization platform, or the game structure (decomposing the attack stages into individual game levels). Since many problems are recurring each semester, the instructors have the solutions ready, which saves the students a lot of time.

\subsubsection{Test run}\label{subsubsec:testrun}
One week before the Open Day, cybersecurity experts review the students' projects and provide feedback so that the students can perfect the game. The test run is an informal live session that takes two hours. The reviewers play the game and immediately discuss its aspects with the student team.

\shiftup
\subsection{Summative Assessment}
The students present their projects to a broad audience after incorporating formative feedback from classmates, tutors, and experts. Finally, the teachers review and mark the final revision of each project. Both occasions are detailed below.

\subsubsection{Open Day}\label{subsubsec:openday}

\begin{figure}
\centering
\includegraphics[width=\linewidth]{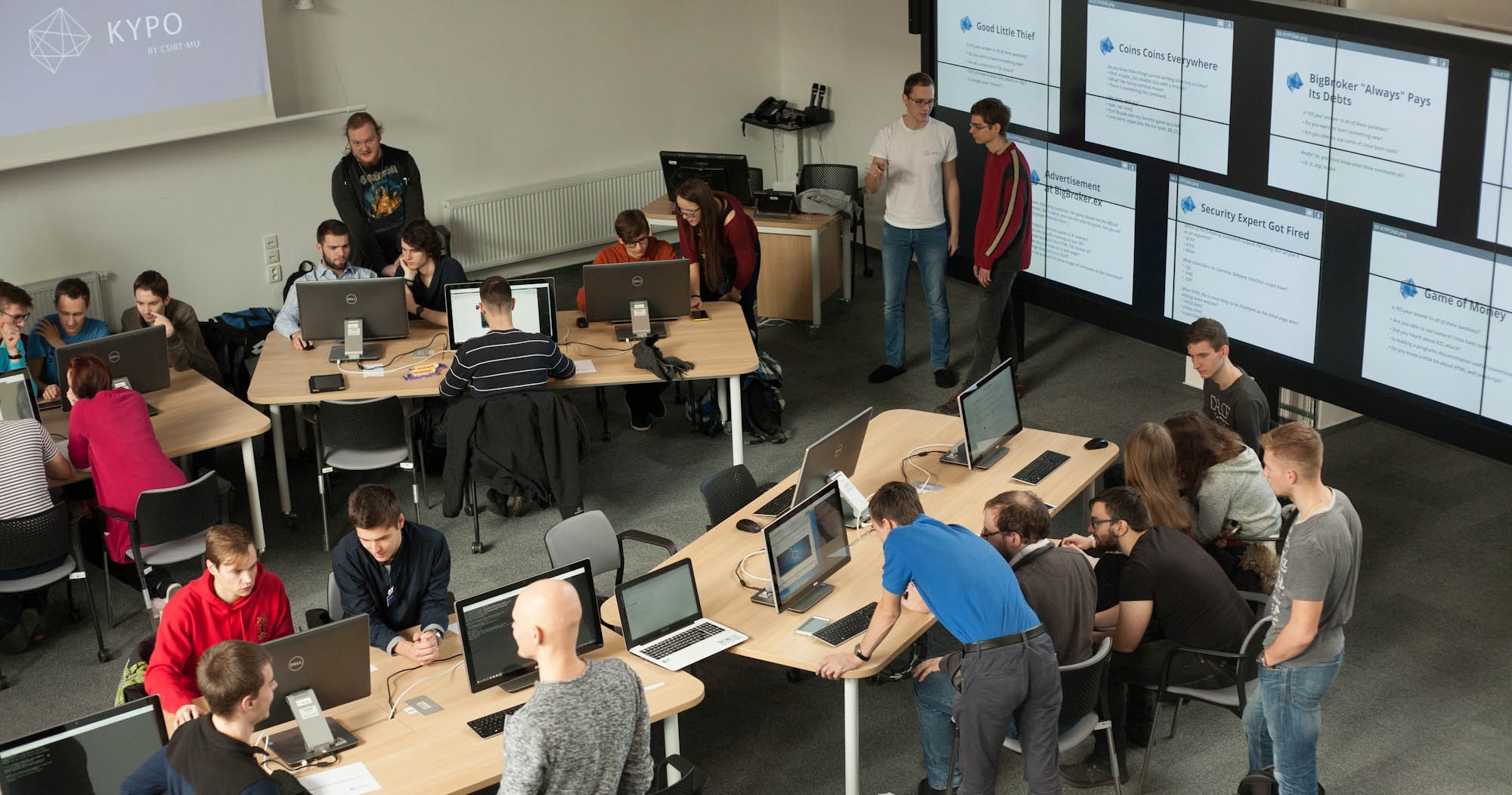}
\caption{Open Day, students present their projects to peers}
\label{fig:open-day}
\shiftup\shiftup
\end{figure}

At the end of the semester, all students of both courses present the games they created on the Open Day event. As Figure~\ref{fig:open-day} shows, the event is very informal, with the goal of promoting the games of our students. During the event, our labs are open to the whole faculty so that anyone can freely come and try any of the students' games.

\subsubsection{Final project review}
After the Open Day, students have the last chance to improve their projects, which they then submit to instructors for a final evaluation. The instructors deploy the projects in the cyber range from scratch to check whether the submission is functional and contains all essential parts, such as documentation of host configuration in the cyber range. Students of the Follow-up course have to use a specific input format that can be processed by a tool for automated installation and configuration, such as Ansible.

\shiftup
\subsection{Student Surveys from the Third Semester}
We now report data and findings from the test run and the Open Day that took place at the end of the third semester, in December 2017. In this semester, we had 18 university students (16 males, 2 females), 15 of which were enrolled in the Introductory course and divided into 7 teams, plus 3 individuals from the Follow-up course.

\subsubsection{Test run}
10 members of the CSIRT volunteered to test the projects. We assigned each tester to one of the 10 student projects with the intention to match the topic of the game and the tester's expertise. Before playing, we informed the testers about the goal of the game, prerequisites, and the estimated time of the gameplay. After playing, the testers completed a survey asking to compare the announced and actual time of playing and rate the educational value of the game, its completeness, and overall quality. The game's educational value was rated on the following scale: None, Small, Medium, High, Huge. The game's degree of completion was rated on a scale from 0 (totally unfinished) to 10 (release candidate). Finally, the game's overall quality was rated on the following scale: Poor, Sufficient, Good, Very good, Excellent.

The announced time for all the games ranged from 15 to 45 minutes, with 30 minutes on average. Among the 10 testers, 4 reported that the estimate was accurate $\pm$ 10 minutes. The remaining 6 of them played longer than estimated, for additional 20 minutes on average. This shows that even in a relatively short game, the students tend to underestimate the required time. Next, 9 out of 10 testers stated that the games have a High educational value, with the possible usage in university or professional education. The completeness ranged from 4 to 8, with the median being 6.5. The overall quality spanned the whole scale, with the median being Good. Finally, there was a very strong Spearman correlation between completeness and quality rating ($\rho = 0.97$, $p<10^{-6}$).

\subsubsection{Open day}
Each of the 7 student teams from the Introductory course was given two computers to run their game; the 3 Follow-up course students had one computer each. Topics of all the games along with the basic prerequisites were displayed on a huge screen 
(see the top right corner of Figure \ref{fig:open-day}). If a visitor selected a game, the student team introduced him/her to the rules and assisted with any difficulties, if the player needed help. We encourage self-reliance in our students, so the teams were fully responsible for attending to the visitors the whole time. Still, the instructors were ready to resolve any technical issues with the cyber range.

After finishing the game, each player reflected on the game in a brief anonymous online questionnaire. The questionnaire was filled in privately on a separate computer so that the players would not be influenced by the presence of the game creators. In the survey, the players rated the educational value of the game and its overall quality on the same 5-step Likert scales we used in the test run. We also measured the total play time of the players. Finally, the players described their learning experiences, so that we could see if the game was perceived as educational, and optionally wrote a subjective comment for the creators, instructors, or organizers.

In total, 41 game plays in teams of one to three people occurred. We report only the aggregate results instead of examining the games separately, due to a small number of participants in each individual game. All of the players provided feedback on the game. The educational value was rated as Medium (9), High (27), and Huge (5). The overall quality was rated as Sufficient (1), Good (10), Very good (23), and Excellent (7). The play time ranged from 5 to 70 minutes, with the median and average being 40 minutes. We attribute the large variance of the time to the fact that some players experienced technical difficulties and had to wait, to different skill levels of the attendees, and to their different game strategies (some just skipped through the game, others wanted to finish it without asking for any hint). We avoid comparing the data of expert reviewers and the attendees, as these two groups perceive the game differently.

The self-reported learning experiences included mostly working with Linux Terminal, using offensive security tools in Kali Linux distribution, and game-specific learning outcomes, such as packet analysis in Wireshark, securing Apache server, or understanding particular vulnerabilities. Of the 41 player teams, 24 included optional comments, which were overwhelmingly positive: 10 students used the word ``super'' in their feedback, 5 especially appreciated how helpful and supportive the student tutors were, and 3 other explicitly asked the organizers to hold more events like this. Only 4 comments were slightly negative: 2 students would have appreciated more precise instructions for playing (that is, what tasks they are supposed to accomplish), and 2 students were bothered by experiencing technical difficulties with the cyber range.

\subsubsection{Limitations of the observations}
The results of the evaluation mentioned above come from a relatively small sample of 18 students. Next, although we encouraged honest feedback, we cannot eliminate the possibility that expert reviewers in the test run might have provided less strict assessment due to student teams being present with them. Finally, the attendees of the Open Day were self-selected, mostly male university students with interest in cybersecurity. While 41 gameplays is a sufficient sample, some of them included groups of two or three people playing one game together (and subsequently, completing the questionnaire together).

\shiftup
\section{Lessons learned} \label{sec:lessons}

This section shares 6 successes and 5 challenges we experienced over the three semesters of teaching and continuous innovation of our courses. These lessons were distilled based on both our observations and feedback of the total of 46 enrolled students, which was gathered by online surveys and informal discussions.

\shiftup
\subsection{Successes}

\subsubsection{A motivating impact of the final presentation}
The Open Day motivates students to work on their project, since they know it will be applied in practice. The students gain an authentic experience of working with a real audience to which they have to present the projects. At the same time, the event poses a strict deadline and pressure.

\subsubsection{Constraining student efforts}
If students are given too much freedom, the complex task of creating a cybersecurity game might be daunting. By specifying constraints such as possible topics, network topology, number of levels, and maximum time, we lowered the barrier for students to start working on the project. Moreover, having a precise specification of the expected result helped the students deliver results of a higher quality. A further restriction on the maximum team size reduced communication overhead and allowed students to focus on the task itself.


\subsubsection{The expert review}
A test run of the scenarios is helpful for the students, as their game is played and reviewed by independent cybersecurity experts. The students can observe how the expert interacts with the game and subsequently improve it. Moreover, the testers can practice skills or enhance knowledge in a different way from their everyday work.

\subsubsection{Recognizing issues early} \label{lessons-early}
Regular checkpoints and especially in-class presentations of the project helped identify and correct students efforts. For example, students who explained the vulnerability exploited in their game to their classmates improved their understanding of it after receiving feedback. However, teams who started with presenting the game proposal realized later that the vulnerability works differently than they thought, which resulted in losing one week of preparation.

\subsubsection{Popularizing cybersecurity at the faculty}
The Open Day builds cybersecurity awareness, substantially promotes our research group and CSIRT, and helps to find new collaborators: especially students for capstone projects, final theses, or further runs of the course. After the third Open Day, we encountered a major increase of requests for thesis supervision or other collaboration.

\subsubsection{Contributing to research and development}
The practical contributions of the courses include developing new serious games applicable for future training, creating opportunities to perform cybersecurity and educational research, and testing of the cyber range. While the platform was being developed, the students and teachers of the courses using it acted as implicit testers, who discovered and reported numerous bugs and feature requests.

\shiftup
\subsection{Challenges}

\subsubsection{Technical infrastructure requirements}
Preparing and employing the cyber range poses an additional burden for both instructors and students. The platform must be in a stable release version and with operational support, as any outages seriously hinder the lessons. We recommend working in the team of at least two lecturers, so that one can fully focus on the content and the other provide technical assistance. In order to relieve teachers of some work, the game creators should have access to the following operations in the cloud: restarting a machine, creating a snapshot, reverting the state of the machine, and editing the game content.

\subsubsection{Preparation and implementation effort from instructors}
Compared to lectures, running hands-on cybersecurity courses introduces a lot of extra work. Apart from technical infrastructure described above, the instructors have to organize the test run and the Open Day, both of which require substantial effort in managing different games real-time on a complex cyber range. What is more, even after the game successfully passes Open Day and final review of the teachers, it still needs further improvement and fine-tuning by experts before it can be used in training sessions.

\subsubsection{Selecting a vulnerability}
The students often select a particular vulnerability for their game that cannot be replicated. The reason is that it has been already fixed in used software, or the vulnerable version is not available anymore, or the current version of the used operating system will not run it anymore. As a result, they are forced to look for a new one and even redesign the game. A possible solution is that instructors would provide a list of vulnerabilities applicable for each selected topic.

\subsubsection{Team formation and teamwork}
As in many other project-oriented courses, the students face the challenge of self-managing a small team. Most of the students report they enjoyed working in a team without experiencing common team issues such as social loafing. However, some students expressed their wish to have more organized and efficient teamwork. Making group processes explicit, for example, by discussing Tuckman stages~\cite{Largent} with the students, can help overcome this issue. Arguably, this would also improve all students in teamwork, since unless group processes are taught explicitly, working on a team project teaches teamwork only implicitly~\cite{isomottonen}. Another difficulty is the grouping into teams itself, which we will explore in our future work.

\subsubsection{Students underestimating complexity of the project}
Creating the game is more complicated than students expect, even though we gradually introduced several checkpoints throughout the semester (see \ref{lessons-early}). Although the students play the exemplary game at the beginning, a brief survey showed that this was not enough for the most to gain a solid understanding of what will involve developing their project. Even when we added extra time for project development at the expense of theory, most teams completed the majority of work only two weeks before the Open Day. Nevertheless, a comment we often hear from students at the end of the course is ``I~wish we started working on this in advance''. Perhaps adding more strict checkpoints mirrored in grading could increase motivation for continuous work during the semester. Right now, we do not grade the course, only give a Pass/Fail mark.

\shiftup
\section{Conclusions} \label{sec:conclusions}

Learning by doing is a popular approach used in cybersecurity education. In alignment with state-of-the-art curricular guidance, we developed two interrelated undergraduate courses that apply learning by teaching in the interactive virtual environment of KYPO cyber range. Students design serious games with the topic of cyber attack or defense, which they then present at the Open Day. They have to cope with numerous interdisciplinary tasks throughout the semester while exercising a broad spectrum of technical and soft skills: system administration, penetration testing, game design, teamwork, project planning, communication, and presentation.

Our experience from three runs of the courses is that they have a strongly beneficial impact on cybersecurity education and research at our faculty. Students rate the courses positively, since they exercise adversary thinking in real-world settings. They can see the practical results of their work during the semester and at the end when presenting their game to their peers. Feedback from attendees of the Open Day shows they enjoyed the event and it attracted some of them to the cybersecurity field. This is highly valuable for security research groups at the faculty, which engages junior collaborators on their research and development projects.

Among the challenges we identified, we suggest prospective instructors who may consider introducing similar courses in their settings, to especially mind the following one. Both courses highly depend on the cyber range hosting the created games. We advise to run them on a stable infrastructure since the topic itself is complex, and every outage of the infrastructure disrupts the sessions. Besides, covering all learning objectives of the Introductory course may lead to shallow instruction in individual areas (network security, system security, game design). The complexity of the cybersecurity field would accommodate two courses, for instance, one covering only the theory and another entirely dedicated to the project.

\shiftup
\begin{acks}
This research was supported by the Security Research Programme of the Czech Republic 2015-2020 (BV III/1--VS) granted by the \grantsponsor{cz-interior}{Ministry of the Interior of the Czech Republic}{http://www.mvcr.cz/bezpecnostni-vyzkum.aspx} under No.~\grantnum{cz-interior}{VI20162019014} -- Simulation, detection, and mitigation of cyber threats endangering critical infrastructure. Martin La\v{s}tovi\v{c}ka is Brno Ph.D. Talent Scholarship Holder -- Funded by the Brno City Municipality.
\end{acks}

\shiftup
\bibliographystyle{ACM-Reference-Format}
\bibliography{references}

\end{document}